\documentclass[aps,twocolumn,showpacs,preprintnumbers,amsmath,amssymb]{revtex4}

\usepackage{graphicx}
\usepackage{dcolumn}
\usepackage{bm}

\begin{document}


\title{Just how different are $SU(2)$ and $SU(3)$ Landau-gauge propagators in the IR regime?}

\author{A. Cucchieri}
\affiliation{Instituto de F\'{\i}sica de S\~aão Carlos, Universidade de
             S\~ao Paulo, \\ Caixa Postal 369, 13560-970 S\~aão Carlos, SP, Brazil}%
\author{T. Mendes}
\affiliation{Instituto de F\'{\i}sica de S\~aão Carlos, Universidade de
             S\~ao  Paulo, \\ Caixa Postal 369, 13560-970 S\~aão Carlos, SP, Brazil}%
\author{O. Oliveira}
\affiliation{Department of Physics, University of Coimbra, 3004 516 Coimbra, Portugal}%
\affiliation{Instituto de F\'{\i}sica de S\~aão Carlos, Universidade de
             S\~ao Paulo, \\ Caixa Postal 369, 13560-970 S\~aão Carlos, SP, Brazil}%
\author{P. J. Silva}
\affiliation{Department of Physics, University of Coimbra, 3004 516 Coimbra, Portugal}%

\date{\today}

\begin{abstract}
The infrared behavior of gluon and ghost propagators in Yang-Mills
theories is of central importance for understanding quark and gluon
confinement in QCD. While simulations of pure $SU(3)$ gauge theory
correspond to the physical case in the limit of infinite quark mass,
the $SU(2)$ case (i.e.\ pure two-color QCD) is usually employed as
a simplification, in the hope that qualitative features be the same
as for the $SU(3)$ case. 
Here we carry out the first comparative study of lattice (Landau) 
propagators for these two gauge groups. Our data were especially produced 
with equivalent lattice parameters in order to allow a careful comparison
of the two cases. We find very good agreement between $SU(2)$
and $SU(3)$ propagators,
showing that in the IR limit the equivalence of the two cases is quantitative, at
least down to about 1 GeV.
Our results suggest that the infrared behavior of
these propagators is independent of the gauge group $SU(N_c)$, as predicted
by Schwinger-Dyson equations.
\end{abstract}

\maketitle

\section{\label{intro}Introduction and Motivation}

Despite recent progress, the infrared structure of Yang-Mills theory is 
still not fully understood. For QCD, the study of the infrared limit is of 
central importance for the comprehension of the mechanisms of quark and gluon 
confinement and of chiral-symmetry breaking. In what concerns confinement, in 
Landau gauge, the infrared behavior of gluon and ghost propagators is linked 
with the Gribov-Zwanziger \cite{Gribov,Zwanziger} and the Kugo-Ojima 
\cite{KugoOjima} confinement scenarios. These confinement mechanisms predict,
at small momenta, an enhanced ghost propagator and a suppression of the gluon
propagator. The strong infrared divergence for the ghost propagator corresponds to a 
long-range interaction in real space, which may be related to quark 
confinement. The suppression of the gluon propagator,
which should vanish at zero momentum, implies (maximal) violation of reflection 
positivity and may be viewed as an indication of gluon confinement. 
Moreover, the interest in the propagators goes beyond the confinement 
mechanism, as they are inputs for many phenomenological calculations in 
hadronic physics (see, for example, Refs.\ \cite{AlkoferSmekal,MarisRoberts}).

Analytic studies of gluon and ghost propagators using Schwinger-Dyson
equations (SDE) \cite{SmekalHA,LercheSmekal,ZwanzigerSDE} seem to agree with the above 
scenarios. (The reader should however be aware that, in the
literature, there are solutions of the SDE \cite{Aguilar03,Aguilar04}
that do not comply with the Gribov-Zwanziger or the Kugo-Ojima
predictions at small momenta.) Moreover, when dynamic quarks are
neglected and assuming that $g^2 \sim 1 / N_c$ --- as suggested by
analysis of the large $N_c$ limit \cite{Hooft} --- the SDE become independent
of the number of color $N_c$. Thus, they predict
that gluon and ghost propagators be independent of $N_c$.

The Landau gauge gluon propagator $D(k^2)$ has been investigated with lattice 
techniques in quenched QCD [i.e.\ pure $SU(3)$ Yang-Mills theory]
\cite{Mandula,Bernard,Marenzoni,Leinweber,Ma,Becirevic,Nakajima,
Bonnet00,Bonnet01,SilvaOliveira04,Furui04,OliveiraSilva05,OliveiraSilva05b,
Sternbeck05,SilvaLat06,SilvaOliveira06,OliveiraSilva06,Sternbeck06,Oliveira:2007dy}, 
in pure $SU(2)$ Yang-Mills theory (in 2, 3 and 4 space-time dimensions)
\cite{Cucchieri97,Cucchieri98,Langfeld,Cucchieri03,Bloch,Cucchieri05,
Cucchieri:2006xi,Cucchieri06,Cucchieri:2006tf,Maas:2006qw,Maas:2007uv} and
in full QCD \cite{Bowman04,Ilgenfritz,Furui06,Bowman07}.
All lattice studies in 4d suggest a finite nonzero infrared gluon propagator
\cite{Bonnet01,OliveiraSilva05b,SilvaOliveira06,Bowman04,Sternbeck06},
in contradiction with the infrared Schwinger-Dyson solution. On the
other hand, finite-size effects are very large and not yet well-controlled,
even in the 3d case \cite{Cucchieri03}. Only in two space-time dimensions
\cite{Maas:2007uv}, using a lattice side $L$ up to about 40 fm, does one find
that $D(0)$ extrapolates to zero as $L$ goes to infinity.
Let us note that investigation of SDE on a 4-torus \cite{Torus} 
suggests that the gluon propagator indeed approaches the
infinite-volume limit very slowly, especially for its low-momentum components.
On the other hand, even with an infrared-finite propagator, one clearly finds
\cite{Langfeld,Cucchieri05,Sternbeck06,SilvaLat06,Bowman07} that 
reflection-positivity
is violated when sufficiently large lattice volumes are considered.
Finally, in the 2d $SU(2)$ case \cite{Maas:2007uv} and in the
4d $SU(3)$ case (using asymmetric lattices) \cite{SilvaOliveira06,Oliveira:2007dy}
it was found that the gluon propagator complies with the pure power-law
behavior predicted analytically \cite{ZwanzigerSDE,SmekalHA}. 

The lattice-Landau-gauge $SU(2)$ and $SU(3)$ ghost propagator 
$G(k^2)$ has been studied in
\cite{Suman,Cucchieri97,Furui04,Boucaud05,Sternbeck05,Boucaud06,Cucchieri05b,
Bogolubsky,Cucchieri06,Cucchieri06c,OliveiraSilva07,Ilgenfritz,Cucchieri:2006tf,
SilvaLat06,OliveiraSilva06,Oliveira:2007dy,Maas:2006qw,Maas:2007uv} and in all cases
an enhancement of the propagator compared to the tree-level behavior $1/k^2$ was observed.
Concerning the comparison between lattice results and the SDE solution, the
two propagators seem to agree only qualitatively. In particular, in three and in four
space-time dimensions, the infrared exponent obtained using lattice simulations
is always smaller than the one predicted analytically. On the other hand,
on the 2d $SU(2)$ case \cite{Maas:2007uv}, the
ghost propagator shows an infrared behavior $1/k^{2.4}$, in agreement with
the SDE solution \cite{ZwanzigerSDE}.

In summary, for the Landau gauge, the SDE gluon and ghost propagators agree, 
at least qualitatively, with the lattice propagators. However, while analytic 
studies using Schwinger-Dyson equations predict {\em the same} infrared behavior
for the $SU(2)$ and $SU(3)$ gauge groups, lattice simulations usually assume 
that the two cases are different, although their qualitative infrared features
may be the same. In this paper, we carry out a comparative study of lattice 
Landau gauge propagators for these two gauge groups. Our data were especially 
produced by considering equivalent lattice parameters in order to allow a
careful comparison of the two cases. We note that we do not assume a power-law
behavior for the propagators, but just compare the raw data in the two
cases.

\section{Numerical Simulations}

\begin{table}
\caption{\label{Tsetup} Lattice setup. The lattice spacing was computed
from the string tension, assuming $\sqrt{\sigma} = 440$ MeV. For $SU(3)$, the
lattice space was taken from \cite{Bali}. The corresponding $\beta$ values
for $SU(2)$ were computed using the asymptotic scaling analysis discussed
in \cite{Bloch}.}
\begin{ruledtabular}
\begin{tabular}{lllll}
 $N^4$    & $a$ (fm)  & $Na$ (fm)  & $\beta_{SU(2)}$  & $\beta_{SU(3)}$ \\
\hline
  $16^4$  &  $0.102$  &  $1.632$  &  $2.4469$  &  $6.0$  \\
  $24^4$  &  $0.073$  &  $1.752$  &  $2.5501$  &  $6.2$  \\
  $32^4$  &  $0.054$  &  $1.728$  &  $2.6408$  &  $6.4$  \\
  $32^4$  &  $0.102$  &  $3.264$  &  $2.4469$  &  $6.0$  \\
\end{tabular}
\end{ruledtabular}
\end{table}

\begin{figure}                                   
\includegraphics[scale=0.33]{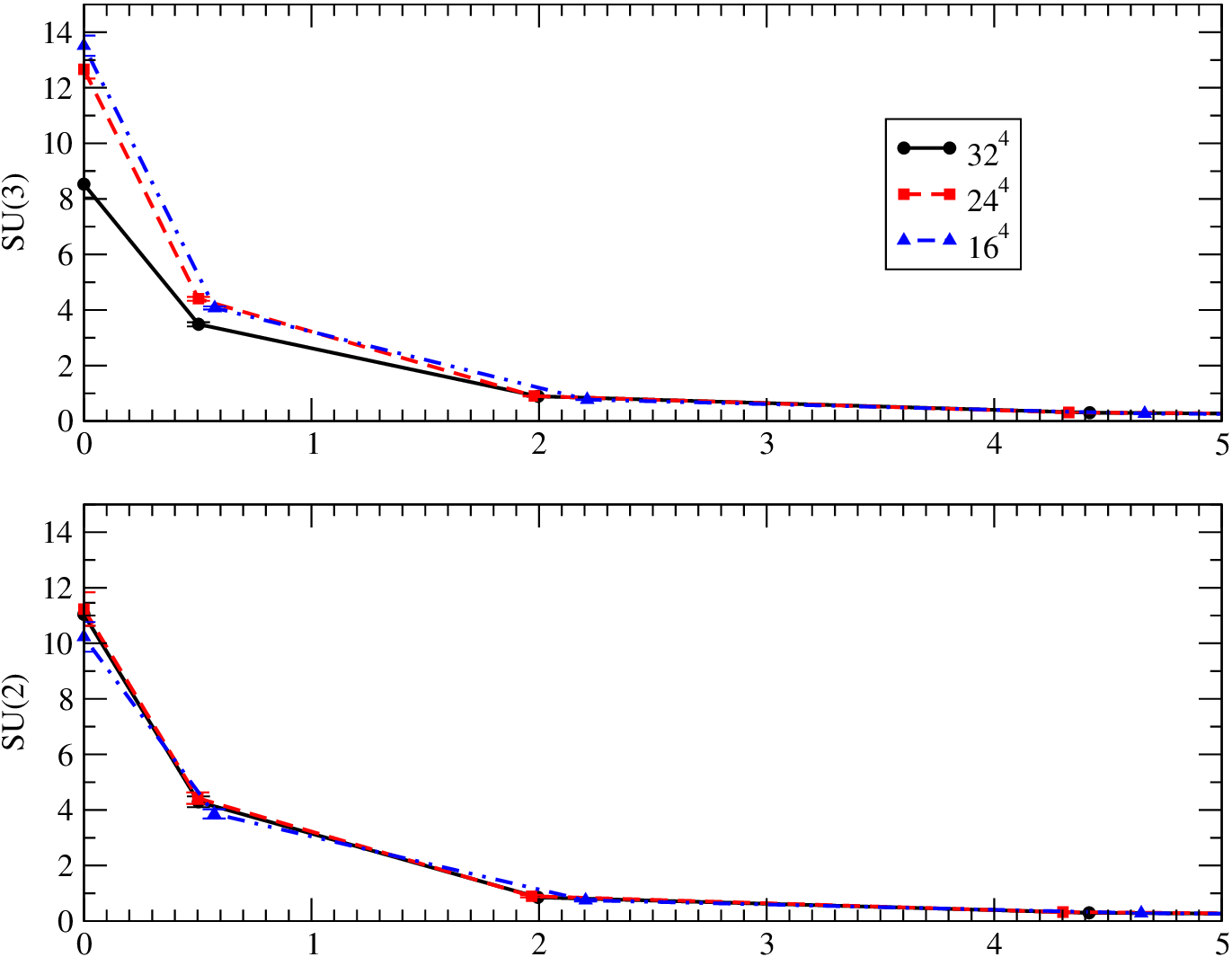}

\vspace{0.63cm}                                   
\includegraphics[scale=0.33]{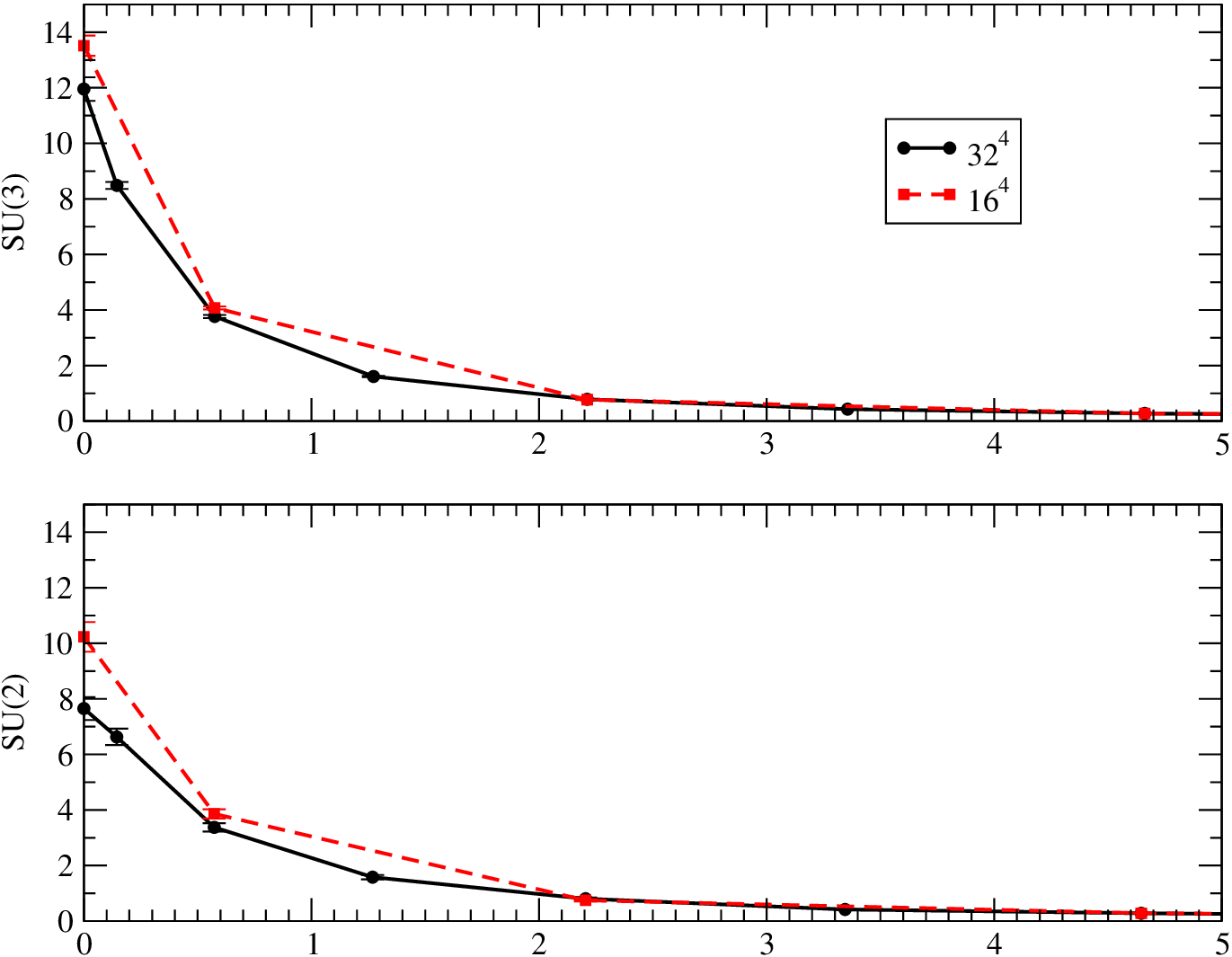}
\caption{\label{gluek000} Renormalized gluon propagator as a function of
the squared magnitude $k^2$ of the four-momentum $k$, for on-axis momenta
$(k,0,0,0)$. We show results
for the $SU(3)$ and the $SU(2)$ cases. In the two top plots we report
data for the three sets of lattice parameters with approximately the
same physical lattice volume $V$. In the two bottom plots we report
data for the two sets of lattice parameters with $\beta = 6.0$ for $SU(3)$
and $\beta = 2.4469$ for $SU(2)$. In all cases we show only data for
$k^2 \leq 5$ GeV$^2$. For larger momenta, the data using different
lattice setups agree well.}
\end{figure}

We consider four different sets of lattice parameters, with the same lattice 
size $N^4$ and the same physical lattice spacing $a$ for the two gauge groups
(see Table \ref{Tsetup}). The first three cases are chosen to yield approximately
the same physical lattice volume $V = (Na)^4 \approx (1.7$ fm$)^4$. This
allows a comparison of discretization effects. The fourth case corresponds to
a significantly larger physical volume, i.e.\ $V \approx (3.2$ fm$)^4$,
in order to study finite-size effects.
For all four cases, 50 configurations were generated~\footnote{The $SU(3)$
configurations were generated with the
MILC code {\it http://physics.indiana.edu/\~\ $\!\!\!$sg/milc.html}.}
using the Wilson action.
The gluon and the ghost propagators
\begin{eqnarray}
   D^{ab}_{\mu\nu} (k^2) ~ = &&
       \delta^{ab} \, \left( \delta_{\mu\nu} - \frac{k_\mu k_\nu}{k^2} \right)
                   \, D( k^2 )  \, , \\
   G^{ab} (k^2) ~ = && - \delta^{ab} \, G( k^2 )
\end{eqnarray}
were computed for four different types of momenta: $(k,0,0,0)$, $(k,k,0,0)$,
$(k,k,k,0)$ and $(k,k,k,k)$. In the computation of $D(k^2)$ and $G(k^2)$, an
average over equivalent momenta and color components was always performed.
In this work we use the field definitions and the choice of momenta reported in
\cite{Bloch} for the $SU(2)$ case and in \cite{SilvaOliveira06} for $SU(3)$.
In particular, each component $k$ is given (in lattice units) by
$k = 2 \sin(\pi n)$, where $n$ is an integer.

\begin{figure}
\includegraphics[scale=0.33]{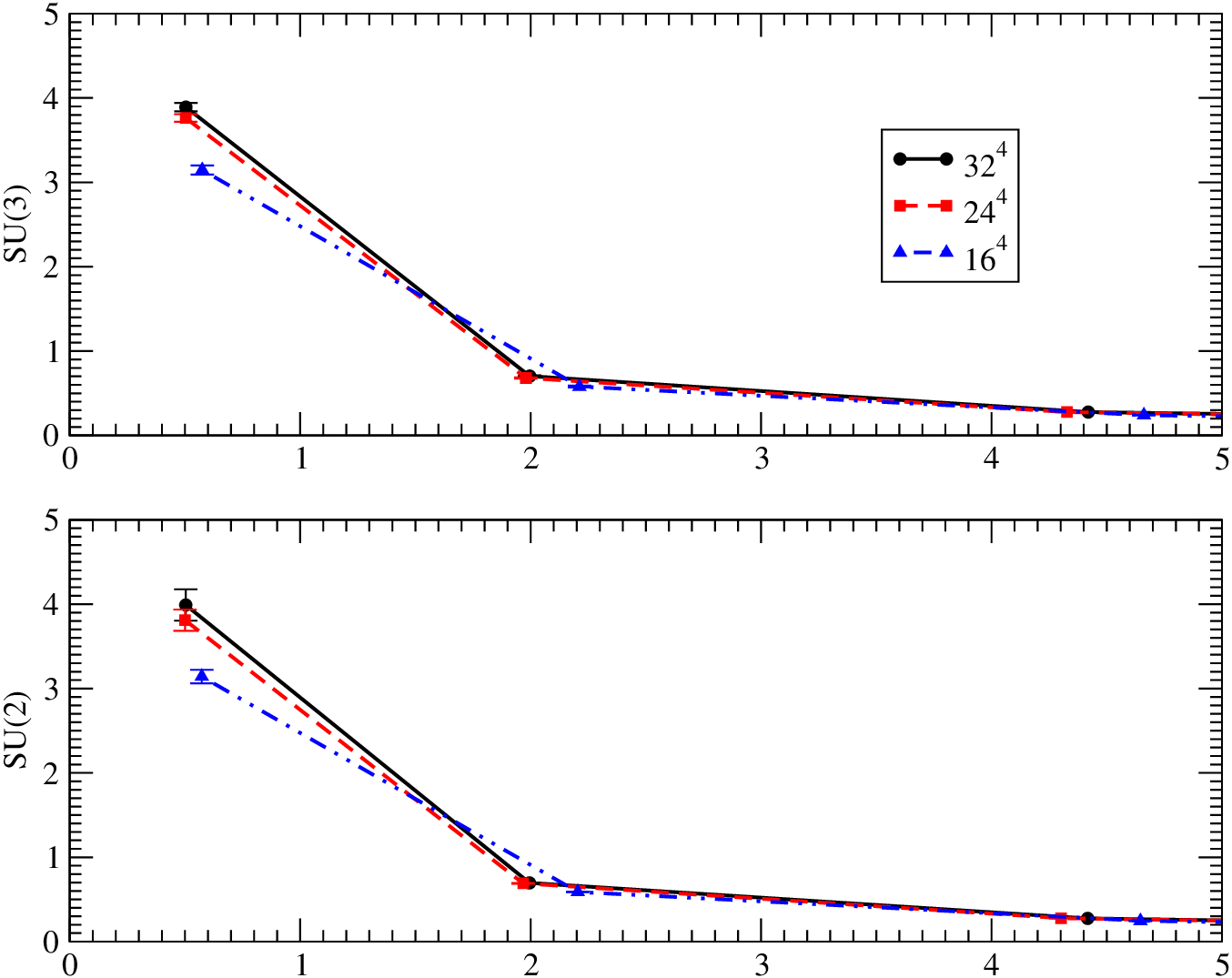}
\vspace{0.57cm}

\includegraphics[scale=0.33]{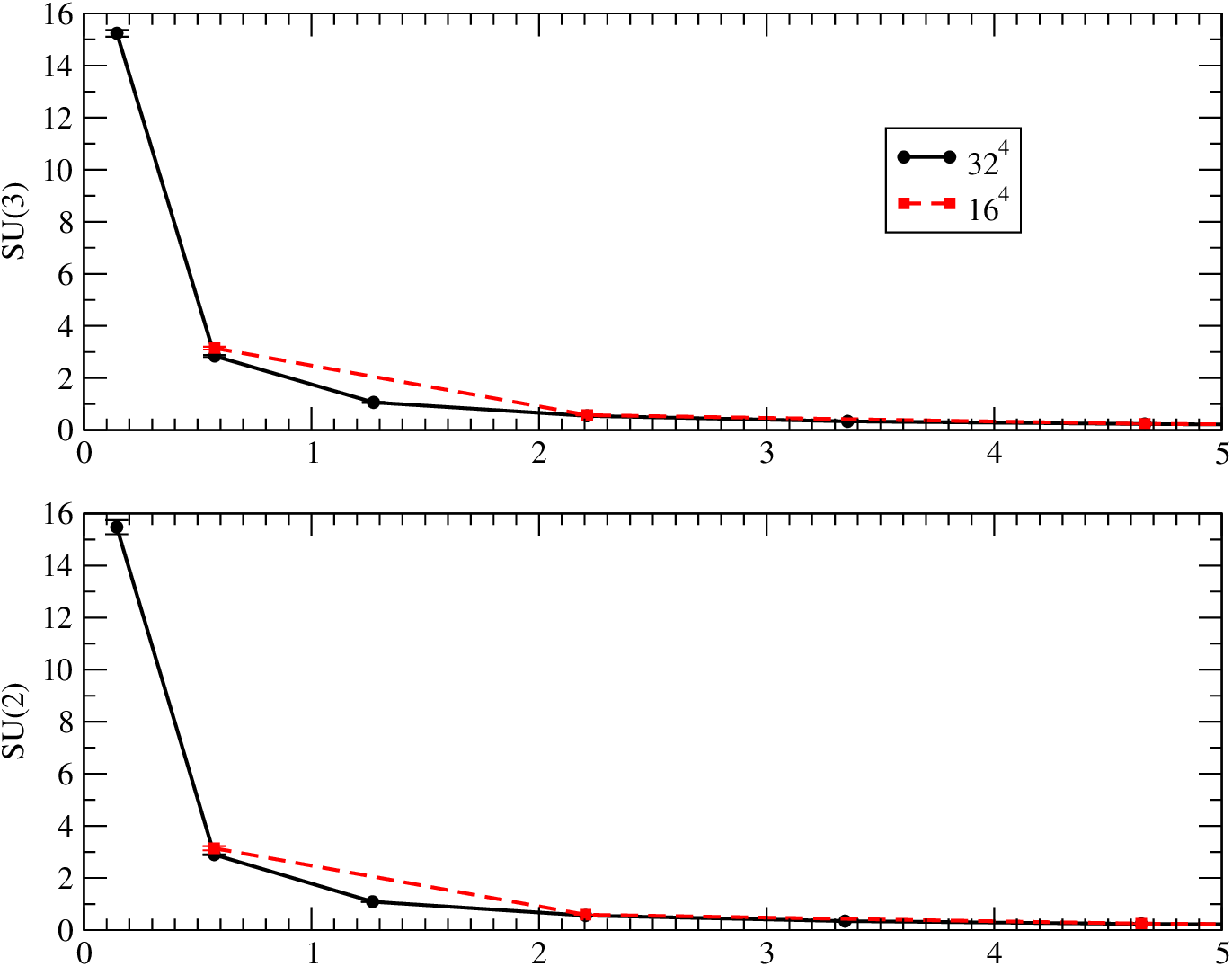}
\caption{\label{ghostk000} Renormalized ghost propagator as a function of
the squared magnitude $k^2$ of the four-momentum $k$, for on-axis momenta
$(k,0,0,0)$.
The data are organized as in Fig.\ \ref{gluek000}.}
\end{figure}

Here, we do not check for possible effects of
the breaking of rotational invariance \cite{de Soto:2007ht}.
In particular, we always compare results for the $SU(2)$
and the $SU(3)$ groups using the same type of momenta in
the two cases.
We also do not consider possible Gribov-copy effects.
Indeed, even though they can play an important role in
the infrared behavior of the propagators \cite{Cucchieri97,
SilvaOliveira04}, with our set of
lattice volumes and for the statistics considered here
these effects should always be smaller than the
statistical error.

The propagators were computed in the minimal Landau gauge, obtained by
minimizing the functional
\begin{equation}
  S[ \Omega ] ~ = ~ - \sum\limits_{x,\mu} \mbox{Tr} \, U^\Omega_\mu (x) \, ,
\end{equation}
where $U^\Omega_\mu (x) = \Omega (x) \, U_\mu (x) \, 
\Omega^\dagger ( x + \hat{e}_\mu)$ is the gauge-transformed link and
$\hat{e}_\mu$ is the unit vector along the $\mu$ direction. For $SU(2)$
the gauge fixing was performed using a stochastic-overrelaxation algorithm
(see \cite{Bloch} for details), while for $SU(3)$ a Fourier-accelerated
steepest-descent algorithm was used (see \cite{SilvaOliveira06} for details).

\begin{figure}
\includegraphics[scale=0.33]{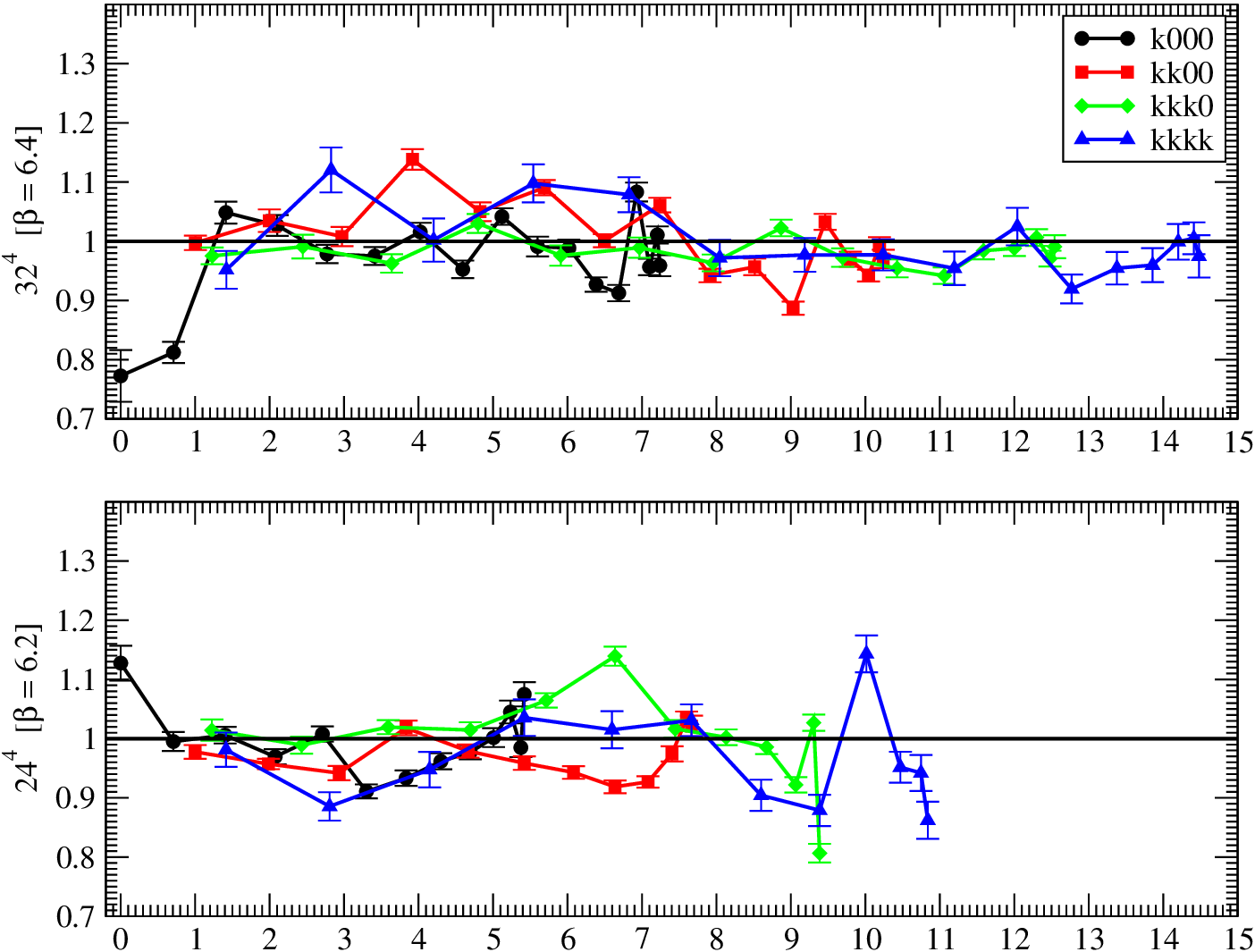}

\vspace{0.63cm}
\includegraphics[scale=0.33]{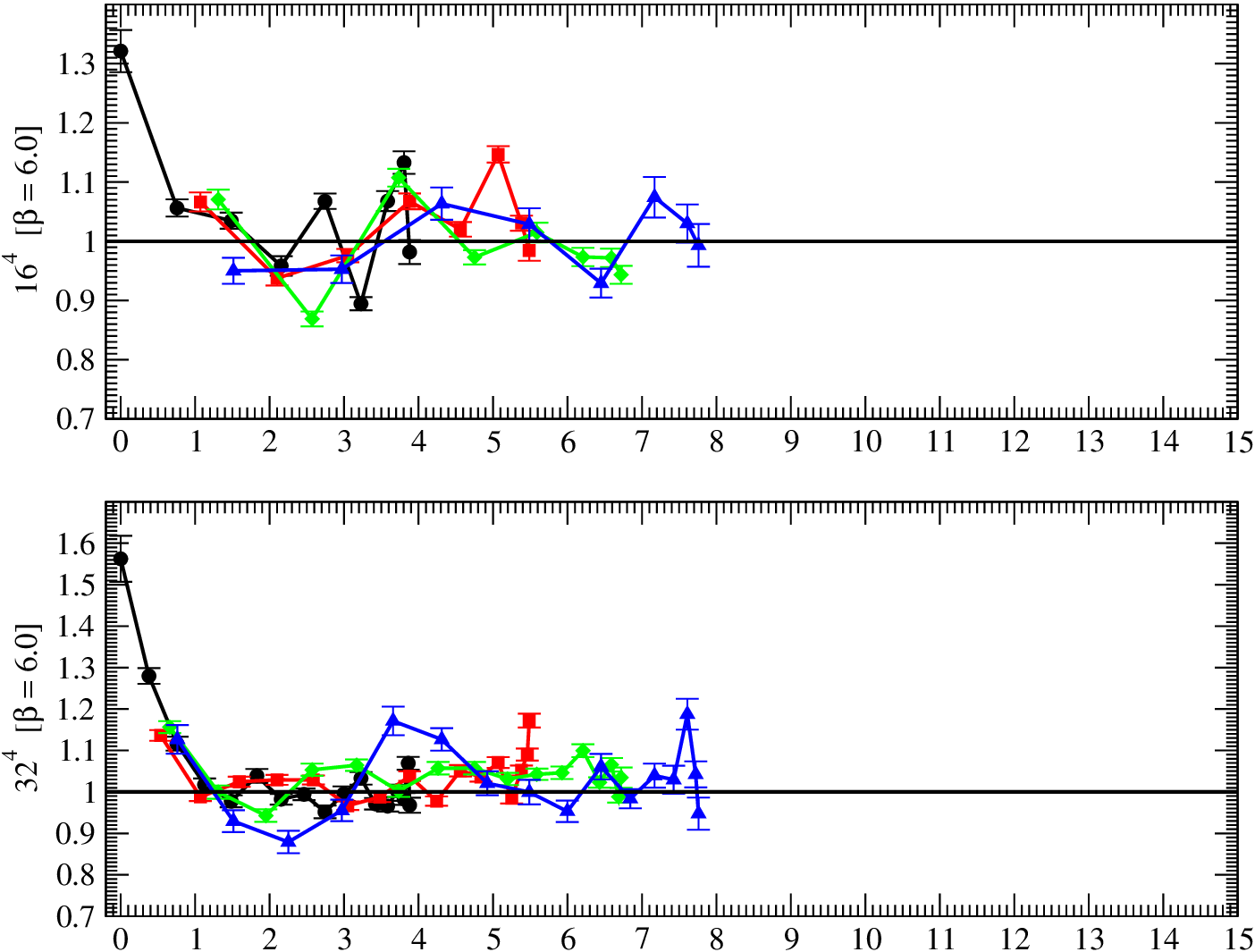}
\caption{\label{ratiosglue} Ratios of $SU(3)$ over $SU(2)$
gluon propagators for the four lattice setups considered.}
\end{figure}

In what concerns the evaluation of the ghost propagator, in the $SU(2)$ case
the Faddeev-Popov matrix was inverted using the method described in \cite{Cucchieri97},
while the $SU(3)$ simulation relies on the method discussed in Ref.\ \cite{Suman}
(considering more than one source). In the calculation of the gluon and of the
ghost propagators, the statistical errors were computed with the
(single-elimination) jackknife method in the $SU(3)$ case and with the
bootstrap method (using 1000 bootstrap samples) in the $SU(2)$ case.
We checked that these errors are in agreement with those obtained considering one
standard deviation.

In order to compare the propagators from the different simulations, the gluon 
and ghost propagators were renormalized accordingly to
\begin{equation}
   \left. D(k^2) \right|_{k^2 = \mu^2} ~ = ~ \frac{1}{\mu^2}, \hspace{1cm} 
   \left. G(k^2) \right|_{k^2 = \mu^2} ~ = ~ \frac{1}{\mu^2},
\end{equation}
using $\mu = 3$ GeV as a renormalization point. The lattice data were
interpolated (using splines) to allow the use of such a renormalization point
in all the simulations. We have checked that the interpolation reproduces perfectly
the lattice data. Let us note that, due to breaking of rotational invariance,
the renormalization factors $Z(\mu^2)$ depend, in general, slightly 
on the type of momenta.
Here we use, for all momenta $k$, the factor $Z(\mu^2)$ obtained from the on-axis momenta
$(k,0,0,0)$.

\section{Results}

\begin{figure}[t]
\includegraphics[scale=0.33]{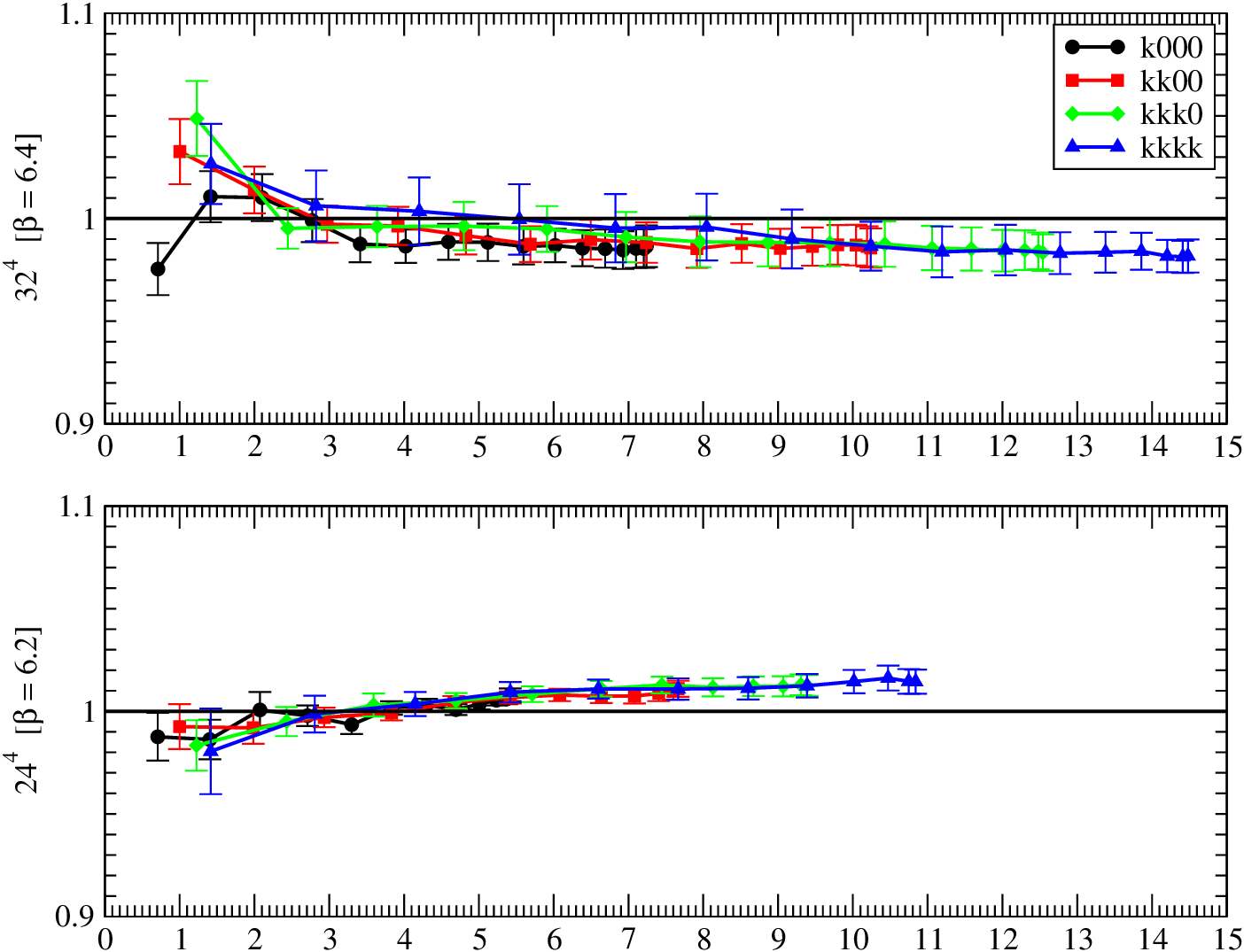}

\vspace{0.63cm}
\includegraphics[scale=0.33]{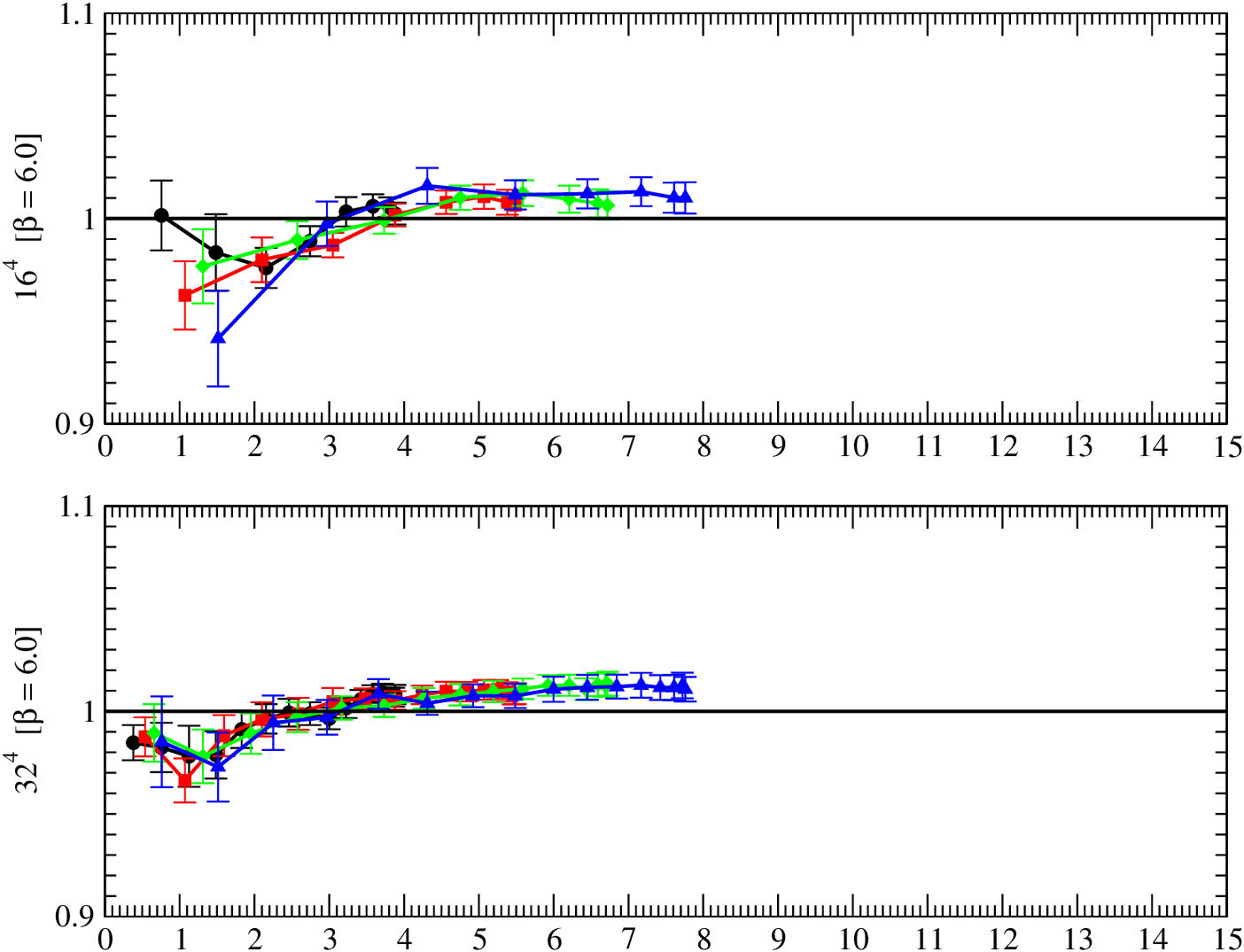}
\caption{\label{ratiosghost} Ratios of $SU(3)$ over $SU(2)$
ghost propagators for the four lattice setups considered.}
\end{figure}

The renormalized $SU(2)$ and $SU(3)$ propagators can be seen
for the various lattice setups in Fig.\ \ref{gluek000} (gluon) and in Fig.\ \ref{ghostk000}
(ghost) for the on-axis momenta $(k,0,0,0)$. (Results are similar when
considering the other types of momenta.) In all figures we report, on the horizontal
axis, the squared magnitude $k^2$ (in GeV$^2$) of the four-momentum $k$. These figures
show that, for the set of momenta accessible in our simulations, finite-volume
and finite-spacing effects are under control. Moreover, they show that the
$SU(2)$ and $SU(3)$ propagators are essentially equal, with slight 
differences in the low-momenta region. Similar results have been recently
presented at {\em Lattice 2007} by Anthony G. Williams \cite{pos340-lat}. In Figs.\
\ref{ratiosglue} (gluon) and \ref{ratiosghost} (ghost) we show the ratios of 
$SU(3)$ over $SU(2)$ propagators. The statistical errors were computed 
assuming Gaussian-error propagation. Note that in the case of the gluon
propagator there are momenta for which the discrepancy from 1 for the ratio
is about 10$\%$ or larger. However, these deviations are not systematic
and are probably due to a combination of several effects. These may include breaking
of rotational invariance, small statistics and finite-size effects, such as
those related to the global $Z(N_c)$ symmetry of the lattice action
\cite{Damm:1998pd,Cucchieri:1999sz,Bogolubsky:2007bw,MM-LAT07}.

\section{Conclusions}

In summary, considering a careful choice of the lattice parameters, 
we were able to carry out an unambiguous comparison of the lattice Landau gluon
and ghost propagators for $SU(2)$ and $SU(3)$ gauge theories. The data
show that the two cases have very similar finite-size and discretization
effects. Moreover, we find very good agreement between the two Yang-Mills 
theories (for our values of momenta larger than 1 GeV),
for all lattice parameters and for all types of momenta.
Below 1 GeV, the results for the two gauge groups show some differences,
especially for the gluon propagator. Note, however, that all ratios are
compatible with 1 within two standard deviations.

In this sense, our results suggest that the propagators are the same
for all $SU(N_c)$ groups in the nonperturbative region, as predicted
by Schwinger-Dyson equations. Of course, given the lattice volumes considered,
further studies are required before drawing final conclusions about the
comparison below 1 GeV. In particular, it will be interesting to investigate
if this agreement persists also in the deep-infrared region, where the gluon
propagator may show a turnover and a suppression, as predicted in the
Gribov-Zwanziger scenario.

\begin{acknowledgments}
The authors thank R. Alkofer, A. Maas and C. Fischer for discussions.
O.O.\ and P.J.S.\ acknowledge FCT for financial support under contract
POCI/FP/63923/2005. P.J.S. acknowledges financial support from FCT via grant
SFRH/BD/10740/2002. O.O. was also supported by FAPESP (grant \# 06/61514-8) 
during his stay at IFSC-USP.
A.C.\ and T.M.\ were supported by FAPESP and by CNPq.
Parts of our simulations have been done on the IBM supercomputer
at S\~ao Paulo University (FAPESP grant \# 04/08928-3) and on the
supercomputer Milipeia at Coimbra University.
\end{acknowledgments}

%

\end{document}